\documentclass[aps,prb,twocolumn,showpacs]{revtex4}

\usepackage{graphicx}
\usepackage{dcolumn}
\usepackage{bm}

\begin{document}

\preprint{APS/123-QED}

\title{Absorption and Emission in quantum dots: Fermi surface effects of Anderson excitons}

\author{R.~W.~Helmes,$^{1}$ M.~Sindel,$^1$ L.~Borda,$^{1,2}$ and J.~von~Delft$^1$}
\affiliation{${}^1$Physics Department, Arnold Sommerfeld Center for Theoretical
Physics, and Center for NanoScience, Ludwig-Maximilians-Universit\"at
M\"unchen, 80333 M\"unchen, Germany \\
$^2$Research Group ``Theory of Condensed Matter'' of the Hungarian Academy of Sciences and Theoretical
Physics Department, TU Budapest, H-1521, Hungary}

\date{February 12, 2005} 

\begin{abstract}
Recent experiments measuring the emission of exciton recombination in a self-organized  
single quantum dot (QD) have revealed that novel effects occur when the wetting layer 
surrounding the QD becomes filled with electrons, because the resulting Fermi sea can hybridize 
with the local electron levels on the dot.
Motivated by these experiments, we study an extended Anderson model, 
which describes a local conduction band level coupled to a Fermi sea, but also includes a local valence band level. 
We are interested, in particular, on how  many-body correlations resulting from the presence of the 
Fermi sea affect the absorption and emission spectra.  
Using Wilson's numerical renormalization group method, we calculate the zero-temperature 
absorption (emission) spectrum of a QD which starts from (ends up in) a strongly 
correlated Kondo ground state.
We predict two features: 
Firstly, we find that the spectrum shows a power law divergence close to the threshold, with an exponent that can be 
understood by analogy to the well-known X-ray edge absorption problem.
Secondly, the threshold energy $\omega_0$ - below which no photon is absorbed
(above which no photon is emitted) - shows a marked, monotonic shift as a 
function of the exciton binding energy $U_{\rm exc}$.

\end{abstract}

\maketitle

\section{\label{Introduction}Introduction}
Recent optical experiments \cite{optical-emission-from2000,hybridization-of-electronic2004} using self-assembled 
InAs quantum dots (QDs), embedded in GaAs, showed that it is feasible to 
measure the absorption and emission spectrum of a single QD.
In absorption spectrum measurements photons are absorbed inside the QD by electron-hole pair (exciton) excitation. 
In emission spectrum measurements, on the other hand, an exciton created by laser 
excitation recombines inside the QD, whereby a photon is emitted which is measured.   

Due to spatial confinement, the QD possesses a charging energy and a discrete energy level structure, 
which can be rigidly shifted with respect to the Fermi energy ${\rm E}_{\rm F}$ by varying an external gate voltage $V_{\rm{g}}$. 
Therefore $V_{\rm g}$ allows for an experimental control of the number of electrons in the QD, 
which in turn determines the energy of the absorbed and emitted photons.
Indeed, the optical data reveal a distinct $V_{\rm g}$-dependence and justify the assumption of a 
discrete energy level structure of the QD \cite{optical-emission-from2000}.


\begin{figure}
\includegraphics[width=1.0\columnwidth]{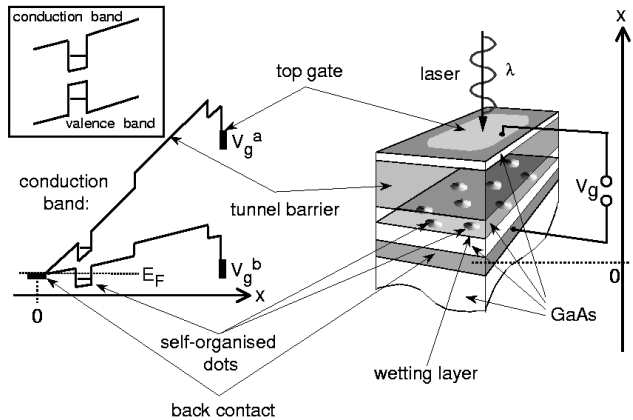}
\caption{
\label{3D-Picture} 
Right hand side: experimental setup used in Refs.\ \cite{optical-emission-from2000,hybridization-of-electronic2004}
(Picture: courtesy of the group of K. Karrai.) 
[bottom to top: GaAs substrate (2000 nm), highly doped GaAs back contact (20 nm, zero-point of x-axis), 
GaAs tunnel barrier (20 nm), InAs mono-layer, forming the wetting layer, together with the QDs, GaAs layer (30 nm), 
AlAs/GaAs tunnel barrier ($\sim$ 100 nm),  GaAs (4nm), NiCr top gate]. 
The gate voltage $V_{\rm g}$, applied between the back contact and the top gate, drives no current through the system, 
since the contacts are separated by two tunnel barriers.  
Left hand side: position-dependence (in $x$-direction) and energy-dependence of the lower conduction 
band edge of the layered structure for two different gate voltages $V^{\rm a}_{\rm g}$ and $V^{\rm b}_{\rm g}$; 
the different band gaps of each material result in jumps at the corresponding interfaces. 
Since the InAs band gap is smaller than that of GaAs, there is a dip in the band gap at the GaAs-InAs interface, 
resulting in QDs with localized conduction and valence band states.
The number of localized electrons trapped in the QD can be controlled by $V_{\rm g}$, 
which shifts the energy levels with respect to the Fermi energy ${\rm E}_{\rm F}$ (set by the back contact).
Inset: holes can be trapped as well due to the bump of the upper band edge of the valence band at the position of the QDs. 
Irradiation by laser light excites electron-hole pairs (excitons) inside the GaAs layer, 
which migrate and become trapped in the InAs QDs.
Finally they recombine by emitting photons, whose emission spectrum is detected.}
\end{figure}

In the experimental set-up, depicted in Fig.\ \ref{3D-Picture}, the InAs QDs are surrounded by an InAs mono-layer, 
called 'wetting layer' (WL), like islands in an ocean.
Above a certain value of $V_{\rm g}$, the conduction band of delocalized states of this 
WL begins to be filled, forming a two-dimensional Fermi sea of delocalized electrons, i.\ e.\ a 
two-dimensional electron gas (2DEG).
The 2DEG hybridizes with localized states of the QD, leading to anomalous emission spectra which could 
not be explained by only considering the discrete level structure of the QD \cite{optical-emission-from2000}.

Motivated by these experiments, we investigate here the optical properties of a QD coupled to a Fermi sea, 
at temperatures sufficiently small that Kondo correlations can occur (T=0).
The Kondo effect in a QD has already been detected in transport 
experiments \cite{kondo-effect-in1998,a-tunable-kondo1998}, where it leads to an enhanced linear conductance.
So far the Kondo effect in QDs has been studied almost exclusively in relation to transport properties.
The experiments of Refs.\ \cite{optical-emission-from2000,hybridization-of-electronic2004} open the 
exciting possibility to study the Kondo effect in optical experiments.

In optics, the effect of Kondo correlations on QDs has to the best of
our knowledge been discussed theoretically only with respect to
non-linear and shake-up processes in a QD
\cite{spin-correlations-in2000, many-particle-resonances2000}.  The
influence of disorder in heavy-fermion systems on the
x-ray-photoemission has been studied e.\ g.\ in 
Ref.\cite{x-ray-photoemission1992}   In this paper we investigate the
absorption and emission spectra of a QD.  We are especially interested
in optical transitions (examples are shown in Fig.\
\ref{StudiedTransition} below) for which the QD starts in or ends up
in a strongly correlated Kondo ground state, and will investigate how
the Kondo correlations affect the observed line shapes.  In Ref.\
\cite{kondo-excitons-in2003} the emission spectrum in the Kondo regime
has already been studied, however with methods which only produce
qualitative results.

The paper is organized as follows: In Section \ref{Model}, we extend the standard 
Anderson model \cite{localized-magnetic-states1961} by including a local valence band level 
(henceforth called v-level) containing the holes.
In contrast to Refs.\ \cite{optical-emission-from2000,hybridization-of-electronic2004} we consider only one 
local conduction band level (henceforth called c-level), to simplify the calculations.
In Section \ref{Method}, we explain how Wilson's numerical renormalization group (NRG) 
method \cite{the-renormalization-group1975} can be adapted to calculate the 
absorption and emission spectrum of the QD.
In Section \ref{Results}, we present the results of our calculations and predict two rather dramatic new features.
Firstly, 
the absorption and emission spectra show a tremendous increase in peak height
as the exciton binding energy $U_{\rm{exc}}$ is increased. 
In fact, the absorption spectrum shows a power law divergence at the threshold energy $\omega_0$, 
in close analogy to the well-known X-ray edge absorption problem\cite{singularities-in-the1969}.
Exploiting analogies to the latter, we propose and numerically verify an analytical expression for the 
exponent that governs this divergence, in terms of the absorption
  (emission)-induced change in the average occupation of the c-level. 
Secondly, the threshold energy below which no photon is absorbed or
above which no photon is emitted, respectively, say $\omega_0$, shows a marked, 
monotonic shift as a function of $U_{\rm{exc}}$; 
we give a qualitative explanation of this behaviour by considering the interplay of 
various relevant energy scales.
Conclusions are given in Section \ref{Conclusions}.

\section{\label{Model} Model}

The experimental setup used in Refs.\ \cite{optical-emission-from2000,hybridization-of-electronic2004}, 
which inspired our analysis, is depicted in Fig.\ \ref{3D-Picture} (see Fig.\ caption for details).
To model this system, we consider an Anderson-like model \cite{localized-magnetic-states1961} for a QD, 
with localized conduction and valence band levels, coupled to a band of delocalized conduction electrons stemming from the WL.
Our model is similar in spirit, if not in detail, to that proposed in 
Refs.\ \cite{kondo-excitons-in2003,coulomb-interactions-in1998}.  
It consists of six terms, illustrated in Fig.\ \ref{ModelHamExplan}:
\begin{figure}
\includegraphics*[width=5 cm]{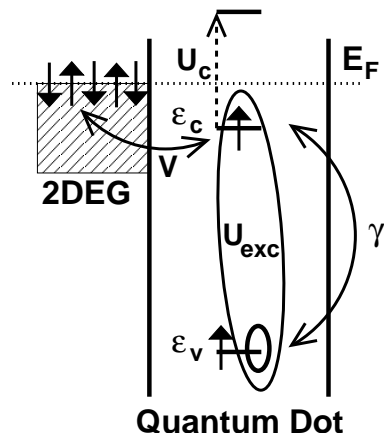}
\caption{\label{ModelHamExplan} 
Model of a semiconductor QD, consisting of one c-level and one v-level, 
with energies $\epsilon_{\rm{c}}$ and $\epsilon_{\rm{v}}$, respectively. 
The Coulomb repulsion of two electrons in the c-level has the strength $U_{\rm c}$. 
The coupling between the c-level and the 2DEG is parametrized by the tunnelling matrix element $V$. 
Crucial for the model is the Coulomb attraction between holes in the v-level and electrons in the c-level, 
which has a strength $U_{\rm{exc}}$. The excitation of electrons from the v-level to the c-level 
(by photon absorption) and the relaxation of electrons from the c-level to the v-level 
(by photon emission) is considered as a perturbation of strength $\gamma$. }
\end{figure}
\begin{equation}
\label{Hamiltonian}
        {\cal{H}} = {\cal{H}}_{0} + {\cal{H}}_{\rm pert},
\end{equation}
where
\begin{equation}
\label{Hamiltonian0}
        {\cal{H}}_{0} = {\cal{H}}_{\rm{c}} + {\cal{H}}_{\rm{v}}
        + {\cal{H}}_{\rm{Uexc}} + {\cal{H}}_{\rm WL} + {\cal{H}}_{\rm c-WL}.
\end{equation}
We consider one c-level with energy $\epsilon_{\rm{c}}$ and one v-level with energy $\epsilon_{\rm{v}}$, 
originating from the conduction or valence band of the InAs QD, respectively.
Note that $\epsilon_{\rm v}$ is smaller than $\epsilon_{\rm c}$ by the order of the band gap; 
since this difference is at least two orders of magnitude larger than all other relevant energy scales, 
its precise value is not important, except for setting the overall scale for the threshold for absorption or emission processes.

Since one c-level is sufficient to produce the effects of present interest, we will, 
in contrast to the experimental situation realized in Refs.\ \cite{optical-emission-from2000,hybridization-of-electronic2004}, 
disregard further local levels to simplify the calculations 
(the experimental situation realized in Refs.\ \cite{optical-emission-from2000,hybridization-of-electronic2004}
will be considered in a future publication).

The c-level and the v-level are described by ${\cal{H}}_{\rm{c}}$ and ${\cal{H}}_{\rm{v}}$, respectively,
\begin{eqnarray}
        {\cal{H}}_{\rm{c}} & = &  \sum_{\sigma} \epsilon_{\rm{c}} \hat{n}_{{\rm c}\sigma}
                + U_c \hat{n}_{{\rm c}\uparrow} \hat{n}_{{\rm c}\downarrow},
\nonumber \\
        {\cal{H}}_{\rm{v}} & = &  \sum_{\sigma} \epsilon_{\rm v} \hat{n}_{{\rm v}\sigma}   
                + U_{\rm v}(1 - \hat{n}_{{\rm v}\uparrow})(1 - \hat{n}_{{\rm v}\downarrow}),
\end{eqnarray}
where $\hat{n}_{{\rm c}\sigma} \equiv c^{\dag}_{\sigma} c_{\sigma}$ and 
$\hat{n}_{{\rm v}\sigma} \equiv v^{\dag}_{\sigma} v_{\sigma}$. Here the Fermi operators $c^{\dag}_{\sigma}$ and 
$v^{\dag}_{\sigma}$ create a spin-$\sigma$ electron in the c-level or in the v-level, respectively. 
The parameters $U_{\rm{c}}$ and $U_{\rm{v}}$ are Coulomb repulsion energies which have to be paid if the c-level 
is occupied by two electrons or if the v-level is empty, respectively.
Since states with two holes are very highly excited states independent of the value of $U_{\rm{v}}$ 
(due to the band gap), the actual value of $U_{\rm{v}}$ has no influence on the results.
The term 
\begin{equation}
\label{Uexc-term}
        {\cal{H}}_{\rm{Uexc}} =  - \sum_{\sigma, \sigma'} U_{\rm{exc}}
                 \hat{n}_{{\rm c}\sigma}(1 - \hat{n}_{{\rm v}\sigma'})
\end{equation}
accounts for the exciton binding energy: 
the Coulomb attraction between each electron in the c-level and each hole in the v-level lowers 
the energy of the system by $U_{\rm{exc}}$. 

The 2DEG formed in the WL is described by 
\begin{equation}
        {\cal{H}}_{\rm{WL}} = \sum_{k, \sigma} \epsilon_{k}
                 l_{k\sigma}^{\dag} l_{k\sigma},
\end{equation}
where the Fermi operator $l_{k\sigma}^{\dag}$ creates a delocalized spin-$\sigma$ electron with wave vector $k$.
The hybridization between the c-level and the 2DEG is described by 
\begin{equation}
        {\cal{H}}_{\rm{c-WL}} = \sum_{k, \sigma}
                V \left(l_{k\sigma}^{\dag} c_{\sigma} 
               +  c_{\sigma}^{\dag} l_{k\sigma} \right),
\end{equation}
where the tunneling matrix element $V$ is assumed to be real and energy-independent.
The hybridization between the c-level and the 2DEG is henceforth parametrized by $\Gamma \equiv \pi \rho_{\rm F} V^2$, 
where $\rho_{\rm F}$ is the density of states (DOS) of the 2DEG at ${\rm E}_{\rm F}$; 
we assume a flat and normalized DOS with bandwidth D.
Since in the considered experiments \cite{optical-emission-from2000,hybridization-of-electronic2004}, 
the mass of the (heavy) holes is significantly larger than the mass of the electrons, we neglect the 
hybridization between the v-level and the 2DEG.

The last part of the Hamiltonian, 
\begin{equation}
        {\tilde {\cal{H}}}_{\rm pert} =  \sum_{k,\sigma} \left(
          \gamma_k
                  \hat{a}_{k} e^{-i \omega_{k} t} c_{\sigma}^{\dag} v_{\sigma} 
		  + \gamma^\ast_k 
\hat{a}^{\dag}_{k} e^{i \omega_{k} t} v_{\sigma}^{\dag} c_{\sigma} \right),
\end{equation}
describing the excitation (first term) and the annihilation (second
term) of excitons in the QD by photon absorption or photon emission,
respectively, is considered as a perturbation of the system.  Here
$\hat{a}_{k}$ ($\hat{a}^{\dag}_{k}$) destroys (creates) a photon of
the laser field with wave vector $k$, where the laser has the
frequency $\omega_{k} = c |k|$.  The coupling is given by $\gamma_k = e
\left( \hbar \omega_{k} / 2 \epsilon_0 V \right)^{\frac{1}{2}}
\vec{\epsilon}_{k} \cdot \vec{D}$, with the elementary charge $e$, the
dielectric constant $\epsilon_0$, the quantization volume $V$, the
orientation of the laser field $\vec{\epsilon}_{k}$, and the dipole
moment of the QD transition $\vec{D}$. Here we assume $\gamma$ to be
independent of $k$, since we consider the laser to be 
approximately monochromatic. Note that for
simplicity, we neglect in the present study terms of the form
$\sum_{k\sigma} \gamma_k \left( l_{k\sigma}^{\dag} v_{\sigma} +
  v_{\sigma}^{\dag} l_{k\sigma} \right)$, describing photon-induced
transitions between the 2DEG and the v-level. Such transitions will
lead to Fano-type effects, which we choose not to consider here, but
will be the subject of future work.  Treating
  ${\tilde{\cal{H}}}_{\rm pert}$ perturbatively is valid as long as
  the optical line width $|\gamma|^2 A$, where $A$ is the density of
  states of the photon field, is small compared to the Kondo
  temperature $T_{\rm K}$ (defined below), the smallest energy scale in our
  studies: $|\gamma|^{2} A \ll T_{\rm K}$.  

In the following considerations, the 
quantized nature of the photon
 field will not play any role in our considerations;
to calculate emission and absorption line
shapes, all that we shall be concerned with are the matrix
elements of the operators $c_\sigma^\dagger v_\sigma +
v^\dagger_\sigma c_\sigma$. For simplicity of
notation, we shall therefore henceforth
write the perturbation term simply  as
\begin{equation}
        {\cal{H}}_{\rm pert} = \gamma  \sum_{\sigma} \left(
                  c_{\sigma}^{\dag} v_{\sigma} + 
v_{\sigma}^{\dag} c_{\sigma} \right).
\end{equation}

\begin{figure}
\includegraphics*[width=1.0\columnwidth]{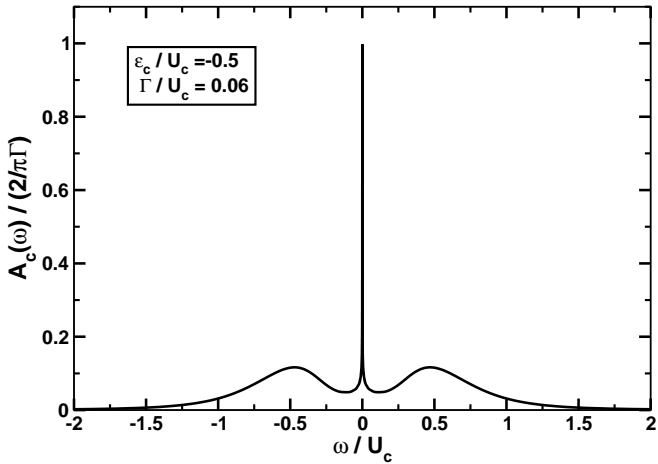}
\caption{\label{Kondoresonance}  
The normalized\cite{friedel-sum-rule1969} local density of states $A_{\rm c}(\omega)$ 
of the c-level in the Kondo regime, with $\epsilon_{\rm c} = - U_{\rm c} / 2$. 
The Kondo effect results in a resonance at the Fermi energy $E_{\rm F}$. 
There are side peaks of the singly (doubly) occupied local level at 
$\omega = \mp U_{\rm c} / 2$ of a level width $2\Gamma$.} 
\end{figure}

  For the scenario of a local spinfull level coupled to a Fermi sea,
  the Kondo effect occurs if the temperature $T < T_{\rm K}$ and the
  average occupancy of the local level is roughly one, i.\ e.\ in our
  case $\langle \hat{n}_{\rm c} \rangle = \sum_{\sigma} \langle
  \hat{n}_{{\rm c}\sigma} \rangle \simeq 1$, known as the 'local
  moment regime' (LMR).  Here $T_{\rm K}$ is given by
\begin{equation}
\label{Kondo-temp}
        T_{\rm K} \equiv \left( U_{\rm c} \Gamma / 2 \right)^{1/2} e^{\pi \epsilon_{\rm c} \left( \epsilon_{\rm c} + U_{\rm c} \right) / 2 \Gamma U_{\rm c}},
\end{equation} 
see Ref.\ \cite{Tsvelick}.  If $T < T_{\rm K}$, $T_{\rm K}$ is the
only relevant energy scale in the problem.  The Kondo effect
introduces a quasi-particle peak, the Kondo resonance, at the Fermi
energy ${\rm E}_{\rm F}$ in the local density of states (LDOS) $A_{\rm
  c}(\omega)$,
\begin{eqnarray}
\label{local-density}
        A_{\rm c}(\omega) & = & \sum_{\tilde {\rm f}, \sigma} 
        \left[ \left| \langle \tilde {\rm f} | c_{\sigma}^{\dag} 
| \tilde {\rm G} \rangle \right|^2 
                 \delta \left(\omega - (E_{\tilde {\rm f}} - 
E_{\tilde {\rm G}}) \right) \right.
\nonumber \\
        & &     \left .+ \left| \langle \tilde {\rm f} 
| c_{\sigma} | \tilde {\rm G} \rangle \right|^2 
                \delta \left(\omega + (E_{\tilde {\rm f}} - 
                  E_{\tilde {\rm G}}) \right)  \right],
\end{eqnarray}
see Fig.\ \ref{Kondoresonance}.  Here $|\tilde {\rm G}\rangle$ and
$|\tilde {\rm f}\rangle$ are eigenstates of ${\cal{H}}_0$ with energy
$E_{\tilde {\rm G}}$ and $E_{\tilde {\rm f}}$, respectively, 
where $|\tilde {\rm G}\rangle$ is
the ground state.  The LDOS $A_{\rm c}(\omega)$ of the c-level is
  a well-known function, which was calculated with the NRG, e. g., by
  Costi et al. \cite{transport-coefficients-of1994}, and has been
  studied frequently since. 

In transport experiments at $T<T_{\rm K}$, the Kondo effect causes the 'zero bias anomaly', 
an enhanced conductance due to the quasi particle peak at $E_{\rm F}$.
Here we will investigate how the Kondo effect affects the absorption and emission spectrum \cite{for-our-numerical-calculations}.

\begin{figure}
\includegraphics*[width=1.0\columnwidth]{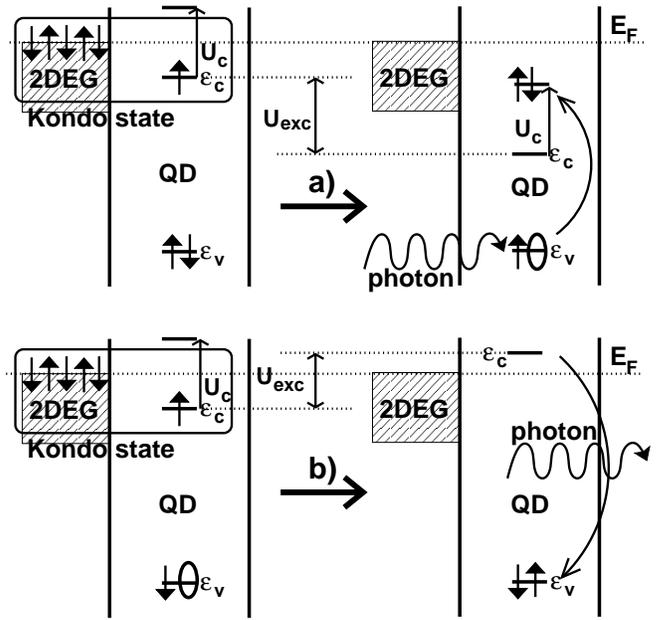}
\caption{\label{StudiedTransition} 
QD (cf. Fig.\ \ref{ModelHamExplan}) tuned such that the c-level is initially (left hand side) singly occupied.
(a) Photon absorption process, inducing a transition from a state 
with no hole in the v-level and a singly occupied c-level 
(Kondo state) to a state with a v-level hole and a doubly occupied c-level (non-Kondo state).
As indicated, the occupation of the c-level in the final state is determined by the value of 
$\epsilon_{\rm c} - U_{\rm exc} + U_{\rm c}$ relative to $E_{\rm F}$. 
(b) Photon emission process, inducing a transition between a state with a v-level hole 
and a singly occupied c-level (Kondo state) to a state without v-level hole and an empty c-level.}
\end{figure}

Fig.\ \ref{StudiedTransition}(a) and Fig.\ \ref{StudiedTransition}(b) show examples of 
absorption and emission processes to be studied in this paper.  
For both examples the QD is tuned such that the c-level is initially singly occupied, 
$\langle \hat{n}_{\rm c} \rangle = 1$, i.\ e.\ in the LMR and therefore gives rise to a strongly 
correlated Kondo state for $T\lesssim T_{\rm K}$.

In the absorption process, Fig.\ \ref{StudiedTransition}(a), a photon excites an electron from the v-level 
into the c-level.
Due to the exciton binding energy, the c-level is 'pulled down' by the value of $U_{\rm{exc}}$. 
Thus the occupation of the c-level in the final state can have any value between one and two, 
depending on the value of $U_{\rm{exc}}$ relative to the charging energy $U_{\rm c}$ of the c-level.
If the final occupation is not in the LMR, the Kondo-state is lost. 

In the emission process, Fig.\ \ref{StudiedTransition}(b), an electron from the c-level 
recombines with a hole in the v-level, thereby emitting a photon. 
In contrast to the absorption process, here the occupation of the c-level decreases since the exciton 
binding energy is lost in the final state. 
Again the Kondo state is lost if the final occupation is not in the LMR.

\section{\label{Method} Method}

The absorption and emission spectra can be calculated using
  Fermi's Golden Rule for the transition rate out of an initial state
  $|{\rm i}\rangle$, which is proportional to
\begin{equation}
\label{Fermis-Golden-Rule-sum}
        \alpha_{\rm i}(\omega) = \frac{2\pi}{|\gamma|^2} \sum_{\rm f}  
        \left|  \langle {\rm f} | {\cal{H}}_{\rm pert} | {\rm i} \rangle \right|^2 
        \delta\left(\omega - \left( E_{\rm f} - E_{\rm i} \right) \right),
\end{equation}
where $|{\rm i} \rangle$ and the possible final states $|{\rm f} \rangle$ are eigenstates of 
${\cal{H}}_0$, cf.\ Eq.\ (\ref{Hamiltonian}), with energy $ E_{\rm i}$ and $E_{\rm f}$, respectively.

No analytical method is known to calculate both the eigenenergies of ${\cal{H}}_0$ and all matrix 
elements $\langle {\rm i} | {\cal{H}}_{\rm pert} |{\rm f} \rangle$ exactly. 
Here we calculate them with Wilson's NRG 
method \cite{the-renormalization-group1975}, a numerically essentially exact method \cite{continuous-spectrum}.

\subsection{\label{NRG} Block structure of Hamiltonian}

Since ${\cal{H}}_0$ commutes with $\hat{n}_{{\rm v}\sigma}$, the number of holes in the v-level is conserved. 
Thus it is convenient to write the unperturbed Hamiltonian ${\cal{H}}_0$ in the basis 
$|\Psi\rangle_{\rm{c+WL}} \otimes |\Psi\rangle_{\rm{v}}$, where $|\Psi\rangle_{\rm{c+WL}}$ 
denotes a product state of the c-level and the 2DEG in the WL, and $|\Psi\rangle_{\rm{v}}$ denotes a state of the v-level.
In this particular basis the unperturbed Hamiltonian ${\cal{H}}_0$ reads
\begin{eqnarray}
\label{block-structure}
        {\cal{H}}_{0} = 
        \bordermatrix{& |0\rangle_{\rm{v}} & |\uparrow\rangle_{\rm{v}} & 
                        |\downarrow\rangle_{\rm{v}} & |\uparrow\downarrow\rangle_{\rm{v}} \cr
                & {\cal{H}}_{{\rm{v}}0} & 0 & 0 & 0 \cr
                & 0 & {\cal{H}}_{{\rm{v}}\uparrow } & 0 & 0 \cr
                & 0 & 0 & {\cal{H}}_{{\rm{v}}\downarrow} & 0 \cr
                & 0 & 0 & 0 & {\cal{H}}_{{\rm{v}}\uparrow\downarrow} \cr},
\end{eqnarray}
where the Hamiltonians
\begin{eqnarray}
\label{block-structure-elements}
        {\cal{H}}_{{\rm{v}}0}   & = &  {\cal{H}}_{\rm{c-WL}} + {\cal{H}}_{\rm{WL}}
                                        + {\cal{H}}_{\rm{c}} 
                                        - \sum_{\sigma} 2 U_{\rm{exc}} \hat{n}_{{\rm{c}}\sigma} + U_{\rm{v}}, 
\nonumber \\
        {\cal{H}}_{{\rm{v}}\uparrow} & = & {\cal{H}}_{\rm{c-WL}} + {\cal{H}}_{\rm{WL}} + 
                                        {\cal{H}}_{\rm{c}} 
                                         - \sum_{\sigma} U_{{\rm{exc}}} \hat{n}_{{\rm{c}}\sigma} + \epsilon_{\rm{v}},
\nonumber \\
        {\cal{H}}_{{\rm{v}}\uparrow\downarrow} & = & {\cal{H}}_{\rm{c-WL}} + {\cal{H}}_{\rm{WL}}
                                        + {\cal{H}}_{\rm{c}} 
                                        + 2 \epsilon_{\rm{v}}
\end{eqnarray}
act only on states $|\Psi\rangle_{\rm{c+WL}}$. 
Since we have not included a magnetic field in our model, ${\cal{H}}_{{\rm v}\uparrow} = {\cal{H}}_{{\rm v}\downarrow}$.

Absorption (A) and emission (E) processes (see Fig.\ \ref{StudiedTransition}) 
involve transitions between different blocks of Eq.\ (\ref{block-structure}):
\begin{eqnarray}
\label{transitions}
  \rm{A:} & & |{\rm G}\rangle = |{\rm G}\rangle_{\rm c+WL} \otimes |{\uparrow\downarrow}\rangle_{\rm v} \rightarrow |{\rm f}\rangle = |{\rm f}\rangle_{\rm c+WL} \otimes |{\sigma}\rangle_{\rm v}
\nonumber \\
  \rm{E:} & & |{\rm g}\rangle = |{\rm g}\rangle_{\rm c+WL} \otimes |\sigma\rangle_{\rm v} \rightarrow |{\rm f}\rangle = |{\rm f}\rangle_{\rm c+WL} \otimes |{\uparrow\downarrow}\rangle_{\rm v}, \nonumber \\ \hspace{0.5 cm}
\end{eqnarray}
where $|{\rm G}\rangle$ is the ground state of ${\cal{H}}_{0}$, $|{\rm G}\rangle_{\rm c+WL}$ 
the corresponding ground state of ${\cal{H}}_{{\rm v}\uparrow\downarrow}$ and $|{\rm g}\rangle_{\rm c+WL}$ 
is the ground state of ${\cal{H}}_{{\rm v}\sigma}$, with $\sigma = \uparrow, \downarrow$.
For absorption, which is governed 
 by $c_{\sigma}^{\dag} v_{\sigma}$, $|{\rm f} \rangle$ 
is a state of the block ${\cal{H}}_{{\rm v}\sigma}$.
For emission, which is governed 
by $v_{\sigma}^{\dag} c_{\sigma}$, 
$|{\rm f} \rangle$ is a state of the block ${\cal{H}}_{{\rm v}\uparrow\downarrow}$.   

To calculate the absorption spectrum, cf.\ Fig.\ \ref{StudiedTransition}(a), 
we insert $|{\rm G}\rangle$ for $|{\rm i} \rangle$ in Eq.\ (\ref{Fermis-Golden-Rule-sum}).
Then $\alpha_{\rm G}(\omega)$ gives the probability per unit time for the transition from $|{\rm G}\rangle$ 
to any final state $|{\rm f} \rangle$ of ${\cal{H}}_{{\rm v}\sigma}$ [containing one hole], 
equivalent to the probability per unit time that a photon with frequency $\omega$ is absorbed, 
which is the desired absorption spectrum $\alpha_{\rm G}(\omega)$, divided by $|\gamma|^2$.
The actual value of $\gamma$ is not important, since it does not affect the shape of the absorption function, but only its height.
The same argument applies to the emission spectrum, cf.\ Fig.\ \ref{StudiedTransition}(b).
Here, one needs to insert $|{\rm g}\rangle$ for $|{\rm i} \rangle$ in Eq.\ (\ref{Fermis-Golden-Rule-sum}).

To employ the NRG to calculate $\alpha_{\rm i}(\omega)$ via Eq.\ (\ref{Fermis-Golden-Rule-sum}), 
one has to overcome a technical problem.
The NRG is a numerical iterative procedure, where the energy spectrum is truncated in each iteration 
[besides the first few iterations].
In standard NRG implementations, transitions from or to highly excited states can only be calculated 
qualitatively rather than quantitatively.
In our case we need to compute transitions to or from states of the blocks ${\cal{H}}_{{\rm v}\sigma}$, 
see Eq.\ (\ref{transitions}), which are highly excited since they are separated by the order of the band 
gap from states of ${\cal{H}}_{{\rm v}\uparrow \downarrow}$, see Section \ref{Model}.
We solve this problem by keeping the same number of states for the blocks 
${\cal{H}}_{{\rm v}\sigma}$ and ${\cal{H}}_{{\rm v}\uparrow\downarrow}$ in each NRG iteration, 
which is in principle the same as running two NRG iterations for both blocks at the same time.
This approach is similar to the one used by Costi et al. \cite{numerical-renormalization-group1994}, 
who studied a problem analogous to ours.

\subsection{\label{LimitingCase} Limiting case of vanishing exction binding energy ($U_{\rm exc}=0$)}

To check the accuracy of the modified NRG method, we begin by considering the limiting case of vanishing exciton binding energy, $U_{\rm{exc}} = 0$.
We will show that for this particular case the absorption and emission spectra are related to the local spectral function.

For $U_{\rm{exc}}=0$ the v-level is decoupled from the c-level and the 2DEG, see Eq.\ (\ref{Uexc-term}).
When decomposing the states in the same way as above, 
$|\Psi\rangle = |\Psi\rangle_{\rm{c+WL}} \otimes |\Psi\rangle_{\rm{v}}$, the total energy can be 
written as a sum, $E = E_{\rm{c+WL}} + E_{\rm{v}}$.
Thus, using Eqs.\ (\ref{Fermis-Golden-Rule-sum}) and (\ref{transitions}), the 
absorption and emission spectrum can be written as
\begin{eqnarray}
\label{Fermis-Golden-Rule-absorption-emission}
        \alpha_{\rm G}(\omega) 
        & = &   2\pi  \sum_{{\rm f}, \sigma} 
                  \left| _{\rm c+WL} \langle {\rm f} | c_{\sigma}^{\dag} | {\rm G} \rangle_{\rm c+WL} \right|^2 \cdot
\nonumber \\
        & &     \times\;\;  \delta\Bigl(\omega - \bigl(E_{{\rm f},{\rm c+WL}} - E_{{\rm G},{\rm c+WL}}\bigl) - 
\Delta \omega\Bigl),
\nonumber \\
        \alpha_{\rm g}(\omega) & = & 2\pi \sum_{{\rm f}, \sigma} 
                  \left| _{\rm c+WL} \langle {\rm f} | c_{\sigma} | {\rm g} \rangle_{\rm c+WL} \right|^2
                \cdot
\nonumber \\
        & &     \times\;\;    \delta\Bigl(\omega - \bigl(E_{\rm f,c+WL} - E_{\rm G,c+WL}\bigl) + \Delta \omega\Bigl).
\nonumber \\  
\end{eqnarray}
Here $\Delta \omega \equiv E_{\rm f,v} - E_{\rm G,v} = - \epsilon_{\rm v}$ represents a constant shift.

To compare the LDOS with the absorption and emission spectrum, we
divide it as $A_{\rm c}(\omega) = A^{+}_{\rm c}(\omega) + A^{-}_{\rm
  c}(\omega) $, with
\begin{eqnarray}
\label{local-density-comparison}
        A^{+}_{\rm c}(\omega) & = & \sum_{f, \sigma} 
                \left| _{\rm c+WL} \langle {\rm f} | c_{\sigma}^{\dag} | {\rm G} \rangle_{\rm c+WL} \right|^2
                \nonumber \\    
        & &  \times\;\;   \delta \Bigl( \omega - \bigl( E_{\rm f,c+WL} - E_{\rm G,c+WL} \bigl)
                 \Bigl)  \hspace{0.2 cm} {\rm for} \hspace{0.2 cm} \omega > 0,
\nonumber \\ 
        A^{-}_{\rm c}(\omega) & = & \sum_{f, \sigma} 
                \left| _{\rm c+WL} \langle {\rm f} | c_{\sigma} |{\rm G} \rangle_{\rm c+WL} \right|^2
                \nonumber \\
        & &  \times\;\;      \delta \Bigl( \omega + \bigl(E_{\rm f,c+WL} - E_{\rm G,c+WL}\bigl)  
                \Bigl) \hspace{0.2 cm} {\rm for} \hspace{0.2 cm} \omega < 0.
\nonumber \\
\end{eqnarray}
Since the operator $c_{\sigma}^{\dag}$ does not change the state of the VB, the sum in 
Eq.\ (\ref{local-density-comparison}) runs only over states $|{\rm f}\rangle$ of ${\cal{H}}_{{\rm v}\uparrow\downarrow}$.

To compare Eqs.\ (\ref{Fermis-Golden-Rule-absorption-emission}) with
Eqs.\ (\ref{local-density-comparison}), note that for $U_{\rm{exc}}=0$
the blocks of the Hamiltonian (\ref{block-structure}) are degenerate
(aside from a constant shift), ${\cal{H}}_{{\rm v}\sigma} =
{\cal{H}}_{{\rm v}\uparrow\downarrow}$, and thus $|{\rm G}
\rangle_{\rm c+WL} = |{\rm g} \rangle_{\rm c+WL}$.  Therefore
\begin{eqnarray}
\label{spectra-comparison}
 \rm{A:} & &  \alpha_{\rm G}(\omega)  =  
2\pi A^{+}_{\rm c}(\omega - \Delta \omega),
\nonumber \\
 \rm{E:} & &  \alpha_{\rm g}(\omega)  =  
2\pi A^{-}_{\rm c}(-\omega - \Delta \omega).
\end{eqnarray}
Thus, for $U_{\rm{exc}}=0$, we can calculate the absorption and
emission spectra in two different ways: firstly, with the modified NRG
procedure and secondly, via Eq.\ (\ref{spectra-comparison}) with
$A_{\rm c}(\omega)$ obtained from the NRG as well.  We find an
excellent agreement between both approaches, which serves as a
consistency check that the modified NRG works as intended.

\begin{figure*}
\includegraphics*[width=12 cm]{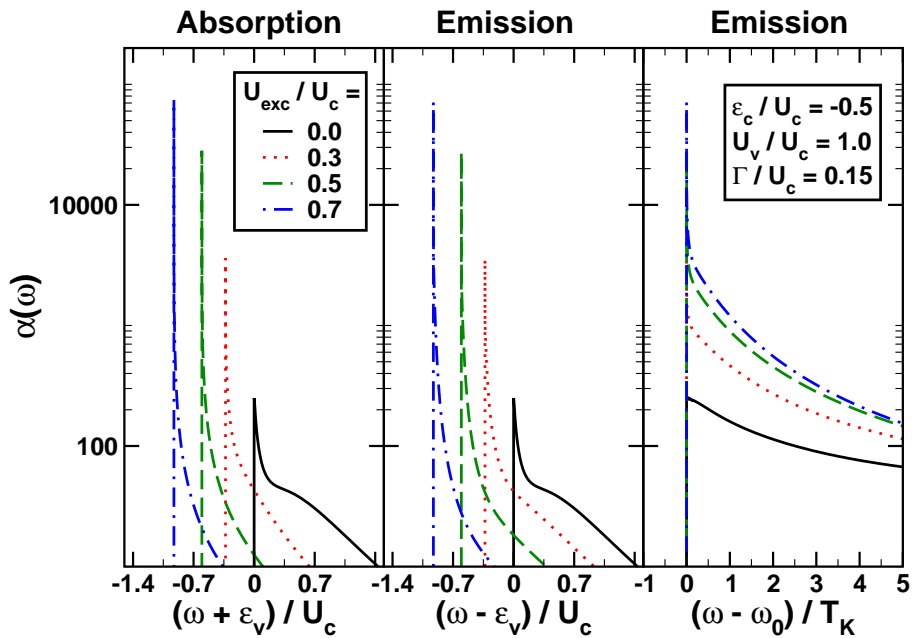}
\caption{\label{AbsandEmi} Absorption and emission spectra for
  different values of $U_{\rm{exc}}$ ($\epsilon_{\rm c}/U_{\rm
    c}=-0.5$,$U_{\rm v}/U_{\rm c}=1.0$ and $\Gamma/U_{\rm c} = 0.15$).
  An increase in $U_{\rm exc}$ results in an increase in the slope of
  the divergence at the threshold-energy $\omega_0$ and in a monotonic
  shift of $\omega_0$ of the absorption and emission spectra. 
The right panel shows the divergence of the emission spectrum at the threshold energy $\omega_0$ 
normalized to $T_{\rm K}$ [$T_{\rm K} / U_{\rm c} = 0.020$, extracted from Eq.\ (\ref{Kondo-temp})].}
\end{figure*}

\section{\label{Results} Results}

In Section \ref{LimitingCase} we showed that for $U_{\rm{exc}}=0$ the absorption or 
emission spectra are related to the LDOS.
Starting from this well-understood limiting case,
let us now study how the absorption and emission spectra behave upon increasing $U_{\rm{exc}}$. 
We use the modified NRG procedure, described in Section \ref{Method}, 
to calculate the absorption and emission spectra $\alpha_{\rm i}(\omega)$ 
from Eq.\ (\ref{Fermis-Golden-Rule-sum}).
The results are shown in Fig. \ref{AbsandEmi}.
We see two striking behaviors: Firstly, there is a tremendous increase in peak height for both 
the absorption and emission spectra. In fact, we find that the spectra diverge at the 
threshold energy $\omega_0$, the energy below which no photon is absorbed or emitted, 
respectively, in close analogy to the well-known 
X-ray edge absorption problem.
Secondly, the threshold energy $\omega_0$ shows a marked, monotonic shift as a 
function of the exciton binding energy $U_{\rm exc}$.

\subsection{Exponent of the power-law divergence}

Let us first study the divergence of the spectral peak at threshold.
For any $U_{\rm exc}\neq 0$, we
find a power-law divergence for both the absorption and the emission spectra\cite{particle-hole},
for energies $\omega$ near the threshold energy $\omega_0$:
\begin{equation}
\label{power-law-divergence}
        \alpha(\omega) \sim \left( \frac{1}{\omega - \omega_0} \right)^{\beta}, \hspace{0.2 cm}\omega \rightarrow \omega_0.
\end{equation}
Examples of this behavior are shown in Fig.\ \ref{log-log-fig}, where
the absorption spectrum is plotted for several different values of
$U_{\rm exc}$ on a double logarithmic plot leading to nice straight
lines for energies $(\omega-\omega_0) < T_K$, i.\ e.\ in the regime
where Kondo correlations can build up.   The slope of such a line
yields the exponent $\beta$.  Remarkably, we find that the exponent so
determined depends only on the change $\Delta n$ in the local
occupation, $\Delta n \equiv \pm \left( \langle n_c \rangle_{\rm
      f} - \langle n_c \rangle_{\rm i} \right)$ ('+' for absorption,
  '$-$' for emission) , to be called 'screening charge', and obeys the
following relation \cite{particle-hole-symmetry-artefact} (who's
origin will be discussed below):
\begin{figure}
\begin{center}
\includegraphics*[width=1.0\columnwidth, angle=0]{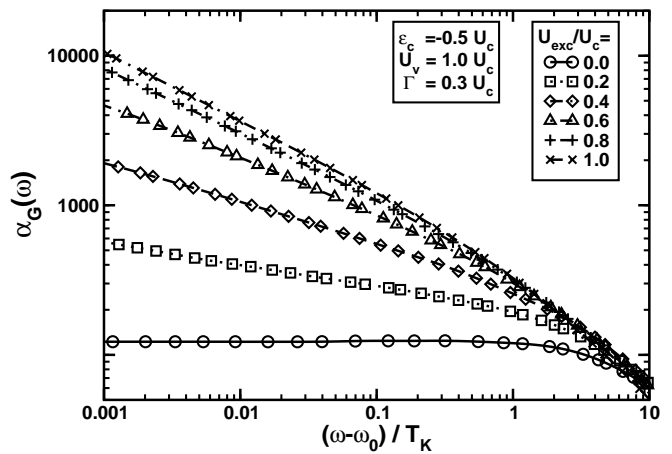}
\end{center}
\caption{Asymptotic behavior ($\omega \rightarrow \omega_0$) of the
  shifted absorption spectra normalized to $T_{\rm K}$ [$T_{\rm
    K}/U_{\rm c} = 0.10$, extracted from Eq.\ (\ref{Kondo-temp})].
  For energies $(\omega - \omega_0) < T_{\rm K}$, in the regime
    where Kondo correlation build up, we find the power-law behavior
    as predicted in Eq.\ (\ref{power-law-divergence}).   The exponent
  $\beta$ increases as $U_{\rm{exc}}$ is increased.  The lower bound
  of $\alpha_{\rm G}(\omega)$ is set by the number of NRG iterations
  (here: $(\omega - \omega_0) \approx 10^{-3} T_{\rm K}$).  The
  asymptotic behavior of the emission spectra (not shown here) looks
  identical to that of the absorption spectra.
\label{log-log-fig}}
\end{figure}
\begin{equation}
\label{exponent-vs-charge2}
        \beta = \Delta n - \frac{(\Delta n)^2}{2}.
\end{equation}
Since for $\omega \rightarrow \omega_0$ the relevant transitions
in the case of absorption and emission are $|{\rm G}\rangle
\rightarrow |{\rm g}\rangle$ and $|{\rm g}\rangle \rightarrow |{\rm
  G}\rangle$, respectively, $\Delta n$ is the same for both types of
transitions, implying the \emph{same} exponent $\beta$ for both
absorption and emission for a given choice of parameters.  In
particular, for the absorption spectra whose asymptotic behavior is
shown in Fig.~\ref{log-log-fig}, we have (for $\omega \rightarrow
\omega_0$) $\langle n_c \rangle_{\rm i}=\langle n_c \rangle_{\rm G}$
and $\langle n_c \rangle_{\rm f}=\langle n_c \rangle_{\rm g}$ where
$\langle n_c \rangle_{\rm g}$ and $\langle n_c \rangle_{\rm G}$
denote the average occupation of the states $| {\rm g} \rangle$ and
$| {\rm G} \rangle$, respectively, see Eq.\ (\ref{transitions}).  At
$U_{\rm exc}=0$, we have $\Delta n = 0$, since there $|{\rm
  G}\rangle_{\rm c+WL} = |{\rm g}\rangle_{\rm c+WL}$, see Section
\ref{Method}.  As $U_{\rm exc}$ increase, $\langle n_c \rangle_{\rm
  g}$ and thus $\Delta n$ also increases, since the Coulomb
attraction between the hole and the electrons in the c-level pulls
down the c-level to an effective value $\tilde{\epsilon}_{\rm c} =
\epsilon_{\rm c} - U_{\rm exc}$ [note that $|{\rm g}\rangle$ is an
  eigenstate of ${\cal{H}}_{{\rm v}\sigma}$, whereas $|{\rm G}\rangle$
  is an eigenstate of ${\cal{H}}_{{\rm v}\uparrow\downarrow}$ and thus
  independent of $U_{\rm exc}$].

\begin{figure}
\begin{center}
\includegraphics*[width=1.0\columnwidth, angle=0]{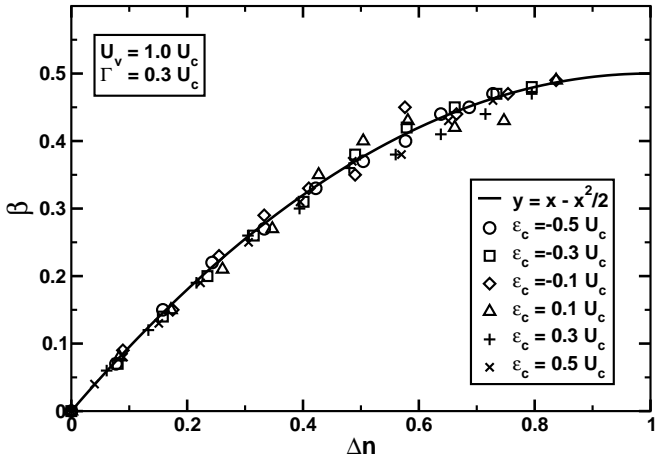}
\end{center}
\caption{ The exponent $\beta$ of the power-law divergence extracted from the NRG results for 
different values of $\epsilon_{\rm c}$ (symbols) coincides very well with the formula 
for $\beta$ given by Eq.\ (\ref{exponent-vs-charge2}) (solid line), indicating that $\beta$ is fully 
determined by $\Delta n$.
Here $\Delta n$ has been varied between $0$ and $\sim 0.8$ by 
varying $U_{\rm exc}$ between $0 U_{\rm c}$ and $1 U_{\rm c}$ in steps of $0.1 U_{\rm c}$.   
\label{exponent-fig}}
\end{figure}

We have extracted the exponent $\beta$ for several different values of $\epsilon_{\rm c}$. 
For each value of $\epsilon_{\rm c}$ we have varied $\Delta n$ between $0$ and $\sim 0.8$ by varying 
$U_{\rm exc}$ between $0$ and $U_{\rm c}$.  
The results are shown in Fig.\ \ref{exponent-fig}.
We find a very good agreement between the results extracted from the NRG and the universal behavior 
predicted in Eq.\ (\ref{exponent-vs-charge2}): all data points nicely collapse onto the curve 
predicted by Eq.\ (\ref{exponent-vs-charge2}).

The numerical results presented in Fig.\ \ref{AbsandEmi} should thus
be interpreted in the following way: for $U_{\rm exc} = 0$ we have
$\Delta n = 0$ and thus $\beta = 0$, which gives a finite height of
the absorption and the emission spectrum at the threshold [in fact the
height is $2 \pi$ times the height of the corresponding LDOS, see Eq.\
(\ref{spectra-comparison})].  As soon as we choose values of $U_{\rm
  exc} > 0$, we find $\beta > 0$, leading to an infinite height of the
absorption and the emission spectral peaks.  Of course, the infinite
peak height is not resolved by our numerical data, for which
$\alpha(\omega-\omega_0)$ is always finite.  However, with increasing
$U_{\rm exc}$ the exponent $\beta$ also increases, resulting in a
steeper slope of the peak at the threshold, which leads to a higher
peak in the numerical results.

\begin{figure*}
\includegraphics*[width=1.0\linewidth]{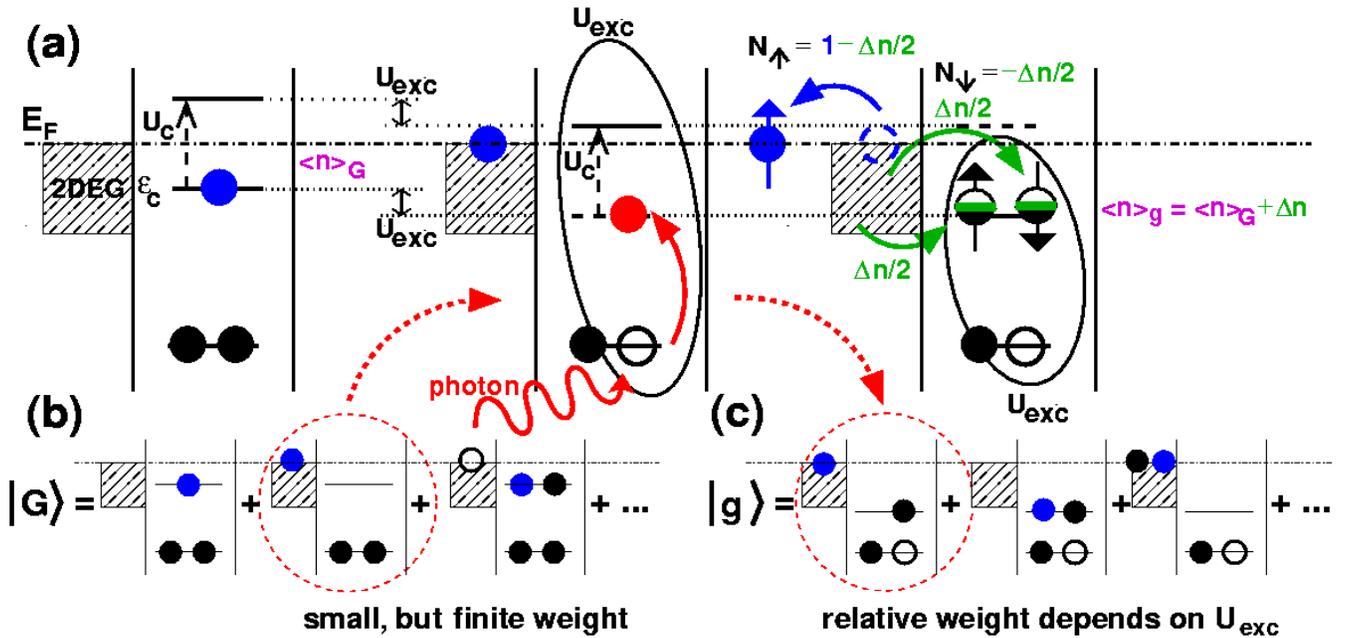}
\caption{\label{Xrayanalog}(a) 
Illustration of an example 
for the absorption process $|{\rm G} \rangle \rightarrow |{\rm g}\rangle$, 
the relevant absorption process for energies $\omega$ close to the threshold $\omega_0$
for the case $\epsilon_c=-U_c/2$, for which the average occupations of the v- and c-levels are 
$\langle n_v \rangle_{\rm{G}}=2$ and $\langle n_c \rangle_{\rm{G}}=1$.
Figs. (b) and (c) schematically depict the  initial and final 
states $|{\rm G}\rangle$ and $|{\rm g}\rangle$, respectively.
Both $|{\rm G}\rangle$ and $|{\rm g}\rangle$ are coherent superpositions of very many different components,
whose c-level can be either empty, singly or doubly occupied;
for $|{\rm G}\rangle$, the v-level is doubly occupied, and the components with empty and  doubly occupied 
c-level have a very small but finite weight.
For $|{\rm g}\rangle$, the v-level is singly occupied,
and since the electron-hole attraction lowers the energy of each c-level electron by $U_{\rm exc}$,
the weight of the components with singly or doubly occupied c-level depends on $U_{\rm exc}$.
In the absorbtion process depicted in (a) and described by the matrix element 
$\langle {\rm g}| c^{\dagger}_{\sigma}v_{\sigma}|{\rm G}\rangle$,
a photon causes the promotion of a spin-$\sigma$ electron from the v-level to the 
c-level (middle panel).
This leads to a transition from a $|{\rm G}\rangle$-component with empty c-level and an extra electron 
in the conduction band (which subsequently flows away from the QD, making a contribution of $+1$
to $N_{\sigma}$), to a state 
with one v-hole and a singly occupied c-level, which in turn is a component of $|{\rm g}\rangle$ [see (c)].
The Coulomb attraction in the state $|{\rm g} \rangle$ between the v-hole and the c-level electrons 
pulls down the c-level from $\epsilon_c$ to an effective value $\tilde{\epsilon}_c = \epsilon_c - U_{\rm{exc}}$,
resulting in an increase of the average c-level occupation $\langle n_{\rm c} \rangle_{\rm g}$  
by $\Delta n$ compared to $\langle n_{\rm c} \rangle_{\rm G}$.
The screening charge $\Delta n=\langle n_c\rangle_{\rm g}-\langle n_c\rangle_{\rm G}$ flows towards the QD, thus 
making a contribution $\Delta n / 2$ to both 
$N_{\sigma}$ and $N_{\bar{\sigma}}$ (with $\bar{\sigma}=\{\downarrow,\uparrow\}$ for $\sigma=\{\uparrow,\downarrow\}$).}
\end{figure*}

An explanation for the universal behavior given by Eq.\ (\ref{exponent-vs-charge2}) can be given by 
studying the analogy between the physics presented in this paper and the well-known X-ray edge absorption problem.
A result analogous to Eq.\ (\ref{power-law-divergence}) was found by Schotte and 
Schotte \cite{threshold-behavior-of1969}, where the absorption spectrum was studied for the X-ray edge problem.
[In Ref.\ \cite{threshold-behavior-of1969} all results are presented for the absorption spectrum. 
However, by rewriting Eq.\ (7) of Ref.\ \cite{threshold-behavior-of1969} for emission, their results 
can be applied to the emission spectrum as well. 
Keeping that in mind, we will focus only on the absorption spectrum in the following, but the argumentation can 
easily be applied to the emission spectrum as well.]  
In Ref.\ \cite{threshold-behavior-of1969},
Fermi-liquid arguments relating phase shifts and local screening charges are used to derive an expression 
for the exponent $\beta$, namely
\begin{equation}
\label{exponent-vs-charge}
        \beta = 1 - \sum_{\sigma} N_{\sigma}^2,
\end{equation}
where $N_{\sigma}$ is the 'effective' number of spin-$\sigma$
electrons [not necessarily an integer] which flow away from the local
level in the absorption process.  Eq.\ (\ref{exponent-vs-charge})
is known as ``Hopfield's Rule of Thumb'' \cite{hopfield1969}.  We can
use this result to analyze our absorption spectra, too, since the
system behaves like a Fermi liquid for $T=0$.  Thus arguments based on
the relation between phase shifts and screening charges do apply.  In
experiments, we expect to find the behavior (\ref{exponent-vs-charge})
for $A |\gamma|^2,T \ll \omega - \omega_0 \ll T_{\rm K}$, where $A
|\gamma|^2$ is the optical line width.

To see that Eqs.\ (\ref{exponent-vs-charge2}) and
(\ref{exponent-vs-charge}) are equivalent, we will now analyze the
absorption process $|{\rm G}\rangle \rightarrow |{\rm g}\rangle$
[relevant absorption process at threshold] and count the charges
$N_{\sigma}$.  It is helpful to consider an example for the process
$|{\rm G}\rangle \rightarrow |{\rm g}\rangle$, shown in Fig.\
\ref{Xrayanalog}, where the initial state $| {\rm G} \rangle$ is the
strongly correlated Kondo ground state with a singly occupied c-level.
The state $| {\rm G} \rangle$ is a coherent superposition of states
with different occupation of the c-level, where the contribution of
the state with empty c-level is small but finite [depicted in Fig.\
\ref{Xrayanalog}(b)].  If the operator $c_{\sigma}^{\dag} v_{\sigma}$
[the part of ${\cal{H}}_{\rm pert}$ corresponding to absorption] is
applied to $|{\rm G} \rangle$, this contribution results in a state
with one hole, a singly occupied c-level and one extra spin-$\sigma$
electron in the conduction band, illustrated in blue in Fig.\
\ref{Xrayanalog}(a).  The latter subsequently flows away from the QD,
making a contribution of $+1$ to $N_{\sigma}$.
This contribution to the final state $c_{\sigma}^{\dag} v_{\sigma} |{\rm G} \rangle$ 
also is a part of the state $|{\rm g} \rangle$, which likewise has contributions from states with empty, 
singly and doubly occupied c-level [Fig.\ \ref{Xrayanalog}(c)]. 
The weight of the contribution with singly occupied c-level to $|{\rm g} \rangle$ depends on $U_{\rm exc}$: 
the Coulomb attraction of the hole in $|{\rm g} \rangle$ pulls down the c-level to the effective value 
$\tilde{\epsilon}_{\rm c}$ resulting in an increase of the average occupation 
$\langle n_{\rm c} \rangle_{\rm g}$ of the c-level by $\Delta n$ compared to $\langle n_{\rm c} \rangle_{\rm G}$, see above.
As $U_{\rm exc}$ is increased, the charge $\Delta n$ [which screens the Coulomb potential of the hole] increases and 
thus the relative weight of the contribution to $|{\rm g}\rangle$ with doubly occupied c-level increases, too, 
whereas the relative weight of the state with singly occupied c-level decreases.
The screening charge $\Delta n$ flows towards the QD,
making equal contributions $-\Delta n/2$ to both $N_{\sigma}$ and $N_{\bar{\sigma}}$.
(with $\bar{\sigma}=\{\downarrow,\uparrow\}$ for $\sigma=\{\uparrow,\downarrow\}$). 
[Another possibility for a transition form $|{\rm G}\rangle$ to $|{\rm g}\rangle$ 
[not depicted in Fig.\ \ref{Xrayanalog}(a)] starts 
from a component of $|{\rm G} \rangle$ with a singly occupied c-level and ends up in a 
contribution of $|{\rm g} \rangle$ with doubly occupied c-level.
One obtains the same results for $N_{\sigma}$ and $N_{\bar{\sigma}}$ if one argues that one unit 
of charge with spin $\sigma$ has to leave the doubly occupied c-level and the charge $\Delta n$ has to flow 
into the c-level to reach the average occupation 
$\langle n_{\rm c} \rangle_{\rm g} = \langle n_{\rm c} \rangle_{\rm G} + \Delta n$.]
Collecting all contributions to $N_{\sigma}$ and $N_{\bar{\sigma}}$, we find
$N_{\sigma}=1-\Delta n/2$ and $N_{\bar{\sigma}}=-\Delta n/2$, which, when inserted into 
(\ref{exponent-vs-charge}), yields Eq.\ (\ref{exponent-vs-charge2}).

A similar argument has been used in \cite{numerical-renormalization-group1994,spectral-properties-of1996}, 
where the local spectral function of the Anderson was studied.


\subsection{Behavior of the threshold energy $\omega_0$}

\begin{figure}
\begin{center}
\includegraphics*[width=1.0\columnwidth, angle=0]{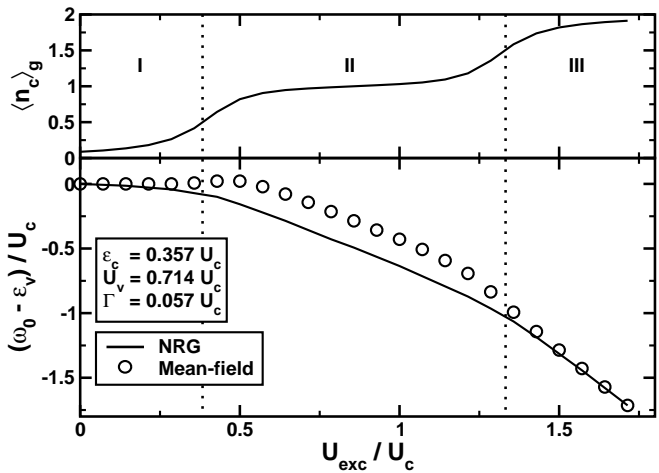}
\end{center}
\caption{Behavior of the threshold energy as a function of $U_{\rm exc}$. 
Upper panel: average occupation $\langle n_{\rm c} \rangle_{\rm g}$ of the c-level for the state
 $|{\rm g}\rangle$, see Eq.\ (\ref{transitions}). 
Three distinct regimes can be identified: empty orbital (I), LMR (II) and full orbital (III)  regime, 
where the c-level is empty, singly or doubly occupied, respectively. 
Since the Coulomb attraction between a hole in the v-level and the electrons in the c-level 
'pulls down' the c-level [$\tilde{\epsilon}_{\rm c} = \epsilon_{\rm c} - U_{\rm exc}$], an increase in $U_{\rm exc}$ 
bears the same effect for $\langle n_{\rm c} \rangle_{\rm g}$ as a decrease in $\epsilon_{\rm c}$.
Lower panel: threshold energy $\omega_0$ versus $U_{\rm{exc}}$ extracted from the NRG results (solid) 
and obtained from the mean-field estimate (circles), Eq.\ (\ref{omega0estimate}), where 
$\langle n_{\rm c} \rangle_{\rm g}$ from the upper panel has been used.
\label{turningpoint1-fig}}
\end{figure}

Let us now consider the second effect observed in Fig.\ \ref{AbsandEmi}, 
the monotonic shift of the threshold energy $\omega_0$. 
The threshold energy for both absorption and emission is given by $\omega_0 = E_{\rm g} - E_{\rm G}$, 
where ${\cal{H}}_0 |{\rm G}\rangle = E_{\rm G} |{\rm G}\rangle$ and 
${\cal{H}}_0 |{\rm g}\rangle = E_{\rm g} |{\rm g}\rangle$, as explained in Section \ref{Method}.
The shift in $\omega_0$ can be understood by considering a mean-field estimate 
of the relevant energies $E_{\rm{G}}$ and $E_{\rm{g}}$:
\begin{eqnarray}
\label{energies-approx}
        E_{\rm{G}} & \simeq & 2\epsilon_{\rm v} + \epsilon_{\rm c} \langle n_{\rm c} \rangle_{\rm{G}}
                        + U_{\rm c} \Big{\langle} \frac{1}{2} n_{\rm c} \Big{\rangle}_{\rm G}^2,
\nonumber \\
        E_{\rm{g}} & \simeq & \epsilon_{\rm v} + \epsilon_{\rm c} \langle n_{\rm c} \rangle_{\rm{g}}
                        +  U_{\rm c} \Big{\langle} \frac{1}{2} n_{\rm c} \Big{\rangle}_{\rm g}^2
                        -U_{\rm{exc}} \langle n_{\rm c} \rangle_{\rm{g}}.
\end{eqnarray}
Here a correlation energy of the order of $T_{\rm K}$ has been neglected.
The average occupations $\langle n_{\rm c} \rangle_{\rm G}$ and $\langle n_{\rm c} \rangle_{\rm g}$ 
can be calculated by NRG. 
Eq.\ (\ref{energies-approx}) allows for a rough estimate of the threshold energy $\omega_0$:
\begin{eqnarray}
\label{omega0estimate}
        \omega_0 & = & E_{\rm{g}} - E_{\rm{G}}  
\nonumber \\
                & \simeq & - \epsilon_v + \epsilon_c \left( \langle n_c \rangle_{\rm{g}} - 
                \langle n_c \rangle_{\rm{G}} \right)
\nonumber \\
                & & + \frac{1}{4} U_c \left(\langle n_c \rangle_{\rm{g}}^2 - \langle n_c \rangle_{\rm{G}}^2 \right)
                - U_{\rm{exc}} \langle n_c \rangle_{\rm{g}}.
\end{eqnarray}
The results for $\omega_0$ shown in Fig.\ \ref{turningpoint1-fig} 
reveal a good agreement between the threshold energy extracted from the absorption and emission spectra 
calculated with NRG (solid line) and the estimation given by Eq.\ (\ref{omega0estimate}).
In the latter approach $\langle n_{\rm c} \rangle_{\rm G}$ and $\langle n_{\rm c} \rangle_{\rm g}$ 
were determined via the NRG, see top panel of Fig.\ \ref{turningpoint1-fig} 
[note that $\langle n_{\rm c} \rangle_{\rm G}$ does not depend on $U_{\rm exc}$ 
and that $\langle n_{\rm c} \rangle_{\rm G} = \langle n_{\rm c} \rangle_{\rm g}$ for $U_{\rm exc}$=0].
We find a linear behavior of $\omega_0$ as a function of $U_{\rm exc}$ for those values of $U_{\rm exc}$ 
where $\langle n_{\rm c} \rangle_{\rm g}$ stays approximately constant.
For this purpose three regions of constant occupation can be identified, 
region I ($\langle n_{\rm c} \rangle_{\rm g} \sim 0$), II ($\langle n_{\rm c} \rangle_{\rm g} \sim 1$) 
and III ($\langle n_{\rm c} \rangle_{\rm g} \sim 2$).
As expected by considering the last term in Eq.\ (\ref{omega0estimate}), 
we observe the slope of $\omega_0(U_{\rm exc})$ to be $0$ in region I, to be $-1$ in region II 
and to be $-2$ in region III, respectively.

The cross-over regions (dotted lines in Fig. \ref{turningpoint1-fig}), 
where $\langle n_{\rm c} \rangle_{\rm g}$ changes between $0$ and $1$ (I $\rightarrow$ II) 
or between $1$ and $2$ (II $\rightarrow$ III), on the other hand, show non-trivial behavior as a function of $\omega_0$.
In these regions the terms in Eq.\ (\ref{turningpoint1-fig}) compete with each other, 
which explains the non-linear behavior (since in these regions $\langle n_{\rm c} \rangle_{\rm g}$ 
itself is a function of $U_{\rm exc}$, too).  

\section{\label{Conclusions} Conclusions}

Motivated by experimental studies of excitons in QDs coupled to a wetting 
layer \cite{optical-emission-from2000,hybridization-of-electronic2004}, 
the aim of this paper was to calculate the absorption and emission spectra of a QD in the strongly 
correlated Kondo ground-state. 
We studied an extended Anderson model, including a local valence band level and a local conduction 
band level which is coupled to a Fermi-sea (2DEG), see Section \ref{Model}. 
For the academic limiting case of a vanishing exciton binding energy,
$U_{\rm exc} =0$, we could relate the absorption and emission spectrum
to the well known local density of states of the local conduction
band, see Section \ref{Method}.  Starting from this limiting case, we
used the NRG to study the spectra for arbitrary values of $U_{\rm
  exc}$.  Our main results are summarized in Fig. \ref{AbsandEmi}
which shows two rather dramatic features: Firstly, an increase in the
slope of the divergence of the absorption and emission spectrum as
$U_{\rm exc}$ is increased.  In fact, the spectra show a power-law
divergence at the threshold energy.
Remarkably, the exponent of the divergence depends only on $\Delta n$, the difference in 
occupation of the local conduction band level between the inital and final states for transitions at the threshold.
We showed that the universal behavior of the exponent can be explained by considering the X-ray edge problem, 
which stands in close analogy to the physics presented in this paper.
Secondly, increasing $U_{\rm exc}$ produces a marked shift of the threshold energy, which can be 
understood rather simply on a mean field level.

In the present paper we considered, for simplicity, a model which contains only a single local 
conduction band level. However, 
in the present generation of experiments \cite{optical-emission-from2000,hybridization-of-electronic2004}, 
the wetting layer forms a 2DEG only for values of the gate voltage $V_{\rm g}$ for which several local
conduction band levels are occupied (not only one, as assumed in the present paper). 
Nevertheless, we expect \cite{private-communications}, that future generations of samples 
could be produced for which the assumptions of our model, namely one c-level with presence of a 2DEG, are fulfilled.
Of course it would be very interesting to generalize our considerations to more general
models, including several local conduction band levels. 

\begin{acknowledgments}
We wish to thank R.\ Bulla, T.\ Costi, A.\ Govorov, A.\ H\"ogele, K.\ Karrai, M.\ Kroner, A.\ Rosch
P. Schmitteckert and S.\ Seidel for helpfull discussions.
This work was supported by the DFG under the SFB 631 and under the CFN, 
'Spintronics' RT Network of the EC RTN2-2001-00440.
L.B. acknowledges support by Hungarian Grants No. OTKA D048665 and
T048782.
\end{acknowledgments}




\bibliography{./Anderson_Excitons}

\begin{thebibliography}{25}
\bibitem[{for()}]{for-our-numerical-calculations}
\bibinfo{note}{For our numerical calculations, we use $|\epsilon_{\rm c}|,
  U_{\rm c} \ll D$, but this is not an essential assumption to realize the
  Kondo effect, which can also occur for $U_{\rm c} \gg D$, as long as $\langle
  \hat{n}_{\rm c} \rangle \simeq 1$.}
\bibitem[{par({\natexlab{a}})}]{particle-hole}
\bibinfo{note}{One might expect that the absorption and emission spectrum are
  related by the particle-hole symmetry. However, the Hamitonian ${\cal{H}}_0$,
  Eq.\ (\ref{Hamiltonian0}), has no particle-hole symmetry: The operators
  $J^{+}\equiv c^{\dag}_{\uparrow}c^{\dag}_{\downarrow}$,
  $J^{-}=(J^{+})^{\dag}$ and $J_{\rm z}=\left[J^{+},J^{-}\right]$ form a SU(2)
  algebra, where $J_{\rm z}=c^{\dag}_{\uparrow}c_{\uparrow} +
  c^{\dag}_{\downarrow}c_{\downarrow}$ is proportional to the charge operator
  and thus commutes with the Hamiltonian. Particle-hole symmetry exists if
  $J^{+}$ commutes with the Hamiltonian. In our case, however,
  $\left[{\cal{H}}_0, J^{+} \right] = (-2\epsilon_{\rm c} - U_{\rm c} + 2
  U_{\rm exc} \sum_{\sigma}(1-n_{{\rm
  v}\sigma}))c^{\dag}_{\uparrow}c^{\dag}_{\downarrow}$. Thus the Hamiltonian
  can only be particle-hole symmetric for one particular choice of $n_{{\rm
  v}\sigma}$, i.\ e.\ different blocks of ${\cal{H}}_0$, see Eq.\
  (\ref{block-structure}) cannot be particle-hole symmetric at the same time.}
\bibitem[{par({\natexlab{b}})}]{particle-hole-symmetry-artefact}
\bibinfo{note}{The fact that the exponent depends only on the local charge of
  the QD is an artefact of the particle-hole symmetry in the WL. In general a
  sum over the charges in the WL will also contribute to the exponent.}
\end{thebibliography}

\end{document}